\documentstyle[preprint,aps]{revtex}
\draft
\tighten

\begin{document}
\date{\today}
\preprint{\vbox{\hbox{hep-ph/9807328}
                \hbox{RUB-TPII-11/98}
                \hbox{WU-B 98-23}}}

\title{  Comment on the paper: ``The perturbative proton form factor
         reexamined'' by Kundu et al.}

\author{ J.~Bolz$^1$, R.~Jakob$^2$, P.~Kroll$^1$
         %\thanks{Email: kroll@theorie.physik.uni-wuppertal.de}
         and N.~G.~Stefanis$^3$
         %\thanks{Email: stefanis@hadron.tp2.ruhr-uni-bochum.de}
       }

\address{{}$^1$Fachbereich Physik,
         Universit\"at Wuppertal,
         D-42097 Wuppertal, Germany \\
         {}$^2$Dipartimento di Fisica Nucleare e Teorica,
         Universit\`{a} di Pavia,
         I-27100 Pavia, Italy \\
         {}$^3$Institut f\"ur Theoretische Physik II,
         Ruhr-Universit\"at Bochum,
         D-44780 Bochum, Germany}

\maketitle
\begin{abstract}
We point out some serious problems in the investigation of Kundu et al.
which claims agreement with the existing data of the proton form
factor, calculated without taking into account the intrinsic 
$k_{\perp}$-dependence of the proton wave function.\\
\end{abstract}

% par1
In 1993 Li \cite{Li93} applied the modified factorization scheme
of perturbative QCD to exclusive processes, developed in \cite{Bo89},
to the proton form factor at large momentum transfer.
One of the key ingredients of his approach was the use of strongly
end-point concentrated distribution amplitudes of the
Chernyak-Zhitnitsky type \cite{CZ84}. With such distribution amplitudes
good agreement of the calculated perturbative contribution to
$G_{M}^{p}$ with the experimental data was claimed.
However, as it was subsequently pointed out in \cite{WUBO}, Li's
analytical expression for the form factor embodied uncontrolled
logarithmic singularities in the soft end-point (kinematic) regions.
As a result, the seemingly good agreement of his numerical results
with experiment was merely the product of an incorrect numerical
integration.

% par 2
Indeed, in \cite{WUBO} we showed in detail where the singularities come
from, and proposed another infrared cut-off prescription which suffices
to render the integrands finite. Our analysis not only comprised gluonic
radiative corrections, encoded in Sudakov form factors as Li's work,
it also took into account the intrinsic $k_{\perp}$-dependence of the
proton wave function. Now this effect is well-known, see, e.g.,
\cite{Col81,Jak93,Sot94,Chi95,Mus97,Akh98}: it has to do with the
finite size of the bound state and has to be taken into account
for consistency. Both the Sudakov form factor and the intrinsic
transverse momentum dependence strongly suppress the end-point
regions,
yielding numerical results for the proton form factor much lower than
the data for a whole class of proton distribution amplitudes, determined
in \cite{BS93}. On the other hand, self-consistency is improved in
our
approach. Furthermore, we recall that our results for the proton form
factor should be considered as a kind of upper bound because
they are obtained under the proviso of normalizing the wave functions
to 1. A more realistic valence Fock state normalization (of the order
of 0.1 to 0.2) leads to substantially smaller perturbative contributions.

% par 3
Now to the present investigation. Kundu, Li, Samuelsson, and Jain
\cite{Kun98} have taken up the calculation of the proton form factor
again, claiming now agreement with experiment for one of the end-point
concentrated distribution amplitudes. The crucial difference
between our investigation \cite{WUBO} and the
new one \cite{Kun98} is the neglect of the intrinsic
$k_{\perp}$-dependence. As we explained above, neglecting this effect
has the consequence of artificially enhancing the perturbative result
at the expense of strong contributions from the end-point regions, where
perturbative QCD is not applicable. There are two more technical
points in which the papers \cite{WUBO}
and \cite{Kun98} differ. One concerns the use of an improved version
of the Sudakov form factor in \cite{Kun98}, which though formally
rectifies the expression given in \cite{Bo89,Li93}, is numerically
irrelevant (see, for instance, the discussion in \cite{Dah95}).
The other point is the use of a larger infrared cut-off in \cite{Kun98},
namely the maximum transverse distance, proposed in our approach
\cite{WUBO}, multiplied by a parameter $c>1$. This cut-off slightly
increases the numerical values of the form factor.

% par 4
Moreover, it is to  be stressed that in the analyses
of \cite{Li93,WUBO,Kun98} the transverse momentum dependence of the
quark propagators is neglected. As discussed in \cite{WUBO} (where we
were mainly interested in estimating the maximum size of the
perturbative contribution) this overestimates the results for the form
factor strongly.

% par 5
Last but not least we want to mention that there are also soft
contributions of the overlap type (Feynman mechanism). Contrary to a
statement made in \cite{Kun98}, these soft contributions do not
constitute an alternative approach to the form factor but are an
essential ingredient of the QCD expansion of the form factors
\cite{Lep80,Rad91,Rad98,Bro98}. Such contributions are
rather to be added to the perturbative contribution.
For the end-point concentrated distribution amplitudes combined with a
plausible assumption on the intrinsic transverse momentum
dependence, these soft contributions are extremely large, overshooting
the data by an order of magnitude \cite{Isg89,Bol96}. Such
distribution amplitudes also lead to inconsistencies with the valence
quark distribution functions \cite{Man89,Bol96}.

To summarize: We still believe that the perturbative contribution to the
proton form factor is much smaller than the experimental data, despite
the claims and the results of \cite{Kun98}, for the reasons explained
above. At accessible momentum-transfer values, the proton form factor
seems to be controlled by soft physics.

%\newpage   %finishes textpart

%\newpage   %finishes references
\end{document}